\documentclass{appolb}
\usepackage{graphicx}
\usepackage{amsmath}
\usepackage{color}
\usepackage{xcolor}
\usepackage{bm}
\usepackage{subcaption}
\usepackage{hyperref}
\usepackage{cite}
\def\cred{\color{black}}

\begin{document}

\title{
\vspace{-4.0cm}
\begin{flushright}
		{\small  LMU-ASC 21/20}   \\
\end{flushright}
\vspace{0.5cm}
From string breaking to quarkonium spectrum%
\thanks{Talk at ``Excited QCD 2020", Krynica Zdr\'oj, Poland, February 2-8, 2020.
}
}
\author{Marco Catillo\thanks{Presenter.},  Marina Krsti\'c Marinkovi\'c
\address{Arnold-Sommerfeld-Center for Theoretical Physics, Ludwig-Maximilians-Universit\"at, Theresienstr. 37, 80333 M\"unchen, Germany}~\\
~\\
{Pedro Bicudo, Nuno Cardoso}
\address{CeFEMA, Instituto Superior T\'ecnico, Av. Rovisco Pais, 1049-001 Lisboa, Portugal}
}

\maketitle
\begin{abstract}
We present a preliminary computation of potentials between two static quarks in $n_f=2$ QCD with O(a) improved Wilson fermions {\cred based on Wilson loops}. We explore different smearing choices (HYP, HYP2 and APE) and their effect on the signal to noise ratio in the computed static potentials. This is a part of a larger effort concerning, at first, a precise computation of the QCD string breaking parameters and their subsequent utilization for the recent approach based on Born-Oppenheimer approximation (Bicudo et al. 2020 \cite{Bicudo:2019ymo}) to study quarkonium resonances and bound states. 
\end{abstract}
  
\section{Introduction}
The computation of quarkonium spectrum is one of the most challenging problems in lattice QCD. 
Recent publications \cite{Bicudo:2019ymo,Prelovsek:2019ywc,Bicudo:2015vta,Bicudo:2016ooe} provided a new and interesting method to study hadron resonances as well as exotic bound states, which are currently posing a challenge for lattice QCD computations. The method is based  on the Born-Oppenheimer approach, which approximates the Hamiltonian for non-relativistic particles, and gives a Schr\"odinger equation that can be  solved numerically. Recent studies \cite{Bicudo:2019ymo,Prelovsek:2019ywc,Bicudo:2015vta,Bicudo:2016ooe} demonstrate promising results regarding energy levels, potentials and wave functions in case where heavy quarks are considered in {\cred the} non-exotic or exotic quarkonium spectrum. One of the necessary ingredients in this approach is the understanding of the string breaking phenomenon \cite{PhysRevLett.81.4056,bali2005observation,koch2019string,Bulava:2019iut,Antonov_2003}, which is commonly described as the breaking of a flux tube formed due to a separation of the heavy quark-antiquark pair. At a large enough distance, the production of light quark-antiquark pairs becomes more favorable than maintaining a flux tube and systems of heavy-light mesons are formed. 
The transition from a quark-antiquark system to a meson-meson system in a $n_f =2$ QCD, can be described by a $2\times2 $ matrix of correlators as outlined in Ref. \cite{bali2005observation}. 
This matrix involves correlators of different operators, namely heavy quark and meson operators, which are not orthogonal and the presence of mixing terms is crucial for the comprehension of such {\cred a} transition. We first focus on the upper left element of such {\cred a} matrix, which is related to the static potential of a quark-antiquark system.

\section{Theoretical aspects}

Given a system of two heavy quarks $Q$ and $\bar{Q}$ with mass $m_Q$ \cite{bali2005observation,PhysRevLett.81.4056,Bulava:2019iut}, the following matrix of correlators:

\begin{equation}
C(t) = \left(\begin{matrix}
C_{QQ}(t) & C_{QB}(t)\\
C_{BQ}(t) & C_{BB}(t)\\
\end{matrix}\right)  
=e^{-2m_Q t}\left(\begin{matrix}
\includegraphics[scale=0.95]{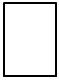} & 
\includegraphics[scale=0.95]{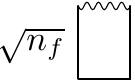} \\
\includegraphics[scale=0.95]{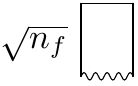} &
\includegraphics[scale=0.95]{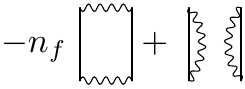} \\
\end{matrix}\right)
\label{eq:bali}
\end{equation}

\noindent
is the crucial tool for studying the string breaking from heavy quarks $Q$ and $\bar{Q}$ to two mesons $B$ and $\bar{B}$. 
The term $C_{QQ}(t)$ represents the correlator of two heavy quarks, $C_{BB}(t)$ is the correlator of the two mesons of the system and $C_{BQ}(t) = C_{QB}(t)$ are the terms denoting the mixing between the physical eigenstates, which are relevant in the transition from a $Q\bar{Q}$ system to a $B\bar{B}$ system.

In this proceeding, we concentrate on the first correlator of Eq. (\ref{eq:bali}), which is basically a Wilson loop $W(r,t)$ up to a prefactor, namely 

\begin{equation}
C_{QQ}(t) = e^{-2m_Q t}W(r,t).
\label{eq:cqq}
\end{equation}

\noindent
From its computation, 
we can obtain the static quark-antiquark potential in the limit of large $t$, i.e.

\begin{equation}
V_{QQ}(r) = \lim_{t\to\infty}\frac{1}{a}\log\left(\frac{C(t)}{C(t+a)}\right) = 
-2m_Q + \frac{1}{a}V(r).
\label{eq:vqq}
\end{equation} 

\noindent
However, for now we do not have access to the parameter $m_Q$, therefore we focus on $V(r)$. 
In fact an additive constant is not relevant in the calculation of physical quantities and we can still fit the potential $V(r)$ with an ansatz 

\begin{equation}
aV_{cont}(r) =  -\frac{\alpha}{r} + c + \sigma r
\label{eq:vr}
\end{equation}

\noindent
where $c$ remains unknown.

\section{\label{sec:tec}Technical aspects}

We consider a set of $79$ CLS\footnote{Coordinated Lattice Simulations, \url{https://wiki-zeuthen.desy.de/CLS/}.} gauge configurations generated with $n_f = 2$ improved Wilson fermions \cite{fritzsch2012strange}.
The lattice parameters are summarized in Table \ref{tab:lattice}, where we have indicated the Sommer parameter as $R_0 = r_0/a$.

\begin{table}[htb]
\centerline{
\begin{tabular}{cccccc}
$V$ & $a$ & $m_{\pi}$ & $R_0$\\
\hline
$32^3 \times 64$ &
$0.0755(11)\,\mbox{fm}$ &
$330 \;\mbox{MeV}$ &
$5.900(24)$ \\
\hline
\end{tabular}}
\caption{Lattice parameters.}
\label{tab:lattice}
\end{table}

These configurations are used to calculate the Wilson loops \footnote{Computed using B. Leder's code (\url{https://github.com/bjoern-leder/wloop/}).},

\begin{equation}
W(r,t)_{lm} = 
\left\langle \Tr\left(U_4(x) U_i(x+t\hat{4})^{(m)}U_4(x+r\hat{i})^{\dagger} U_i(x)^{(l)\;\dagger} \right)\right\rangle,
\label{eq:wl}
\end{equation}

\noindent
where $U_{\mu}(x)$ is a generic Wilson line at the point $x$ in direction $\mu$. 
$W(r,t)$ is the main ingredient for the computation of the static potential as showed in Eqs. (\ref{eq:cqq}) and (\ref{eq:vqq}). 
In Eq. (\ref{eq:wl}) the labels $l$ and $m$ refer to different smearing levels, which are only applied on the Wilson lines in the spatial direction.
The amount of smearing can be represented as an operator $\mathcal{S}_{sm}$, namely $U_{\mu}(x)^{(l)} = (\mathcal{S}_{sm})^{n_l}U_{\mu}(x)$.
In our study every configuration is, at first, smeared with either HYP or HYP2 smearing. 
The difference of {\cred these two} is in the choice of the smearing parameters, i.e. HYP: $\alpha_1 = 0.75,\; \alpha_2 = 0.6, \;\alpha_3 = 0.3$, and HYP2: $\alpha_1 = 1.0,\; \alpha_2 = 1.0, \;\alpha_3 = 0.5 $;
see Refs. \cite{hasenfratz2001flavor,donnellan2011determination} for their meaning.
As {\cred a} next step we apply further smearing of the spatial links (we call it as sHYP smearing) as defined in Eq. (\ref{eq:wl}).
The sHYP smearing is chosen with parameters: $ \alpha_2 = 0.6, \;\alpha_3 = 0.3$ (namely only two smearing steps are applied, according to Ref. \cite{donnellan2011determination}), where we take the following smearing levels: $0,4,10$. 
We also consider an APE smearing (always in the spatial direction) with two choices {\cred for the} parameter $\alpha$, i.e. $\alpha = 0.5\;\mbox{or} \;\alpha = 0.7$.
Furthermore we study the generalized eigenvalue problem (GEVP), solving the system $\hat{W}(r,t) \bm{v}  =\lambda(r,t)\hat{W}(r,t_0)\bm{v}$, 
with $\hat{W}(r,t) = \left(  W(r,t)_{lm}\right) $, where $t_0  =a$ is kept fixed. 
Then the ground state potential is extracted knowing that
\begin{equation}
V(r) = \lim_{t\to\infty}\frac{1}{a}\log\left(\frac{\lambda(r,t)}{\lambda(r,t+a)}\right),
\label{eq:vqq2}
\end{equation} 
\noindent
where $\lambda(r,t)$ is the largest eigenvalue of the GEVP. 
\noindent
The matrix $\hat{W}(r,t)$ is chosen to be a $5\times 5$ matrix with different sHYP smearing  levels ($ \alpha_2 = 0.6, \;\alpha_3 = 0.3$) in the spatial direction, namely $n_l=0,3, 6, 9, 12$; which are chosen according to the formula $n_l = (l/12)R_0^2$, see Ref. \cite{donnellan2011determination}.
Finally we compare the described smearing choices with {\cred the} case where no smearing is applied. 
The aim was to observe how sensitive  {\cred our data are} to different smearing strategies and we have chosen some of the most commonly used techniques in  {\cred the} literature. 

\section{Results}

\begin{figure}[htb]
\centerline{
\includegraphics[scale=1.1]{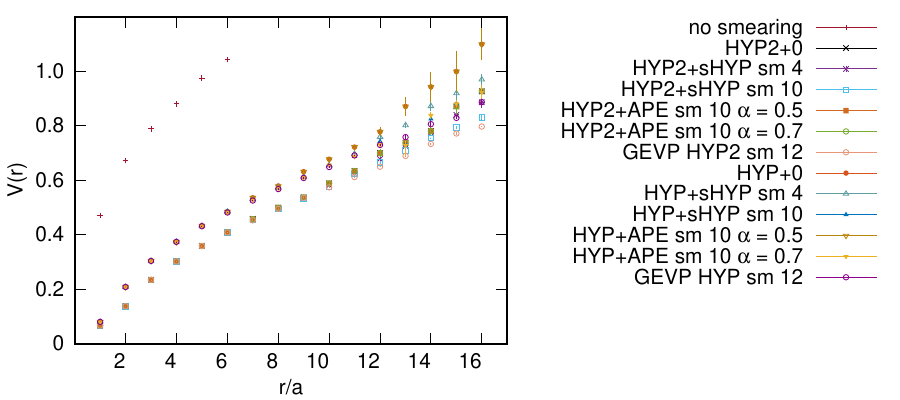}}
\caption{Potentials for different HYP2 and HYP smearing and the no-smearing case.}
\label{fig:pot-all}
\end{figure}

\begin{figure}[htb]
\centering
\begin{subfigure}{0.49\textwidth}
\centering
\includegraphics[scale=1.1]{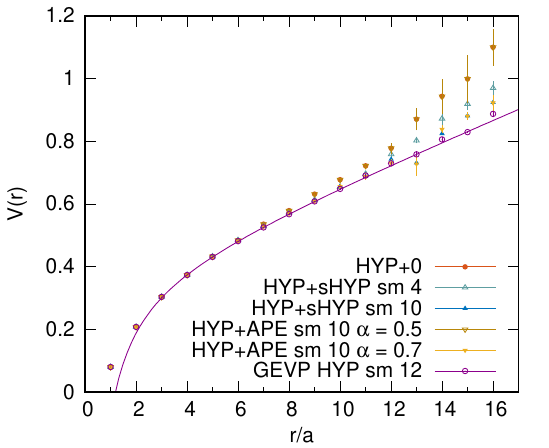}
\end{subfigure}
\begin{subfigure}{0.49\textwidth}
\centering
\includegraphics[scale=1.1]{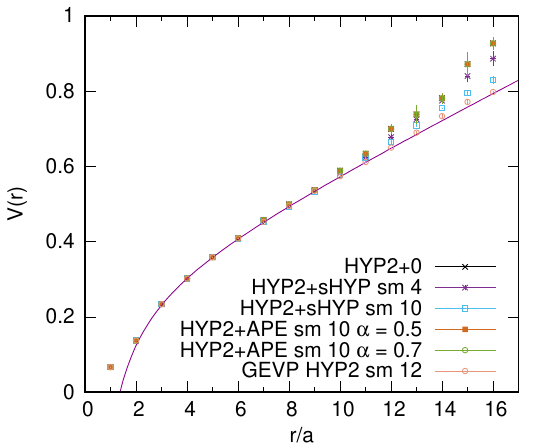}
\end{subfigure}
\caption{Potentials for different HYP smearing (left) and HYP2 smearing (right). The fit curves are for the case of GEVP with HYP and HYP2 smearing with $5$ levels of smearing $n_l=0,3,6,9,12$, see section \ref{sec:tec}.}
\label{fig:pot-HYP}
\end{figure}

In Fig. \ref{fig:pot-all}, we present the results for the potential $V(r)$ for different smearing choices, where the jackknife method is used for the first estimate of the errors.\footnote{N. Cardoso's code \texttt{qfit} is used for analysis (https://github.com/nmrcardoso/qfit).}
We observe that {\cred in the case without smearing}, we only get a few points for small $r$, since for large $r$ the signal to noise ratio deteriorates and the plateau cannot be reliably determined. 
However, already one level of smearing (HYP or HYP2) is enough to get an acceptable signal and plot $V(r)$ for all $r$. 
Furthermore, the curve with HYP smearing is shifted up with respect the HYP2 smearing and this is evident for the range $r/a \in [2,11]$, where the signal is more clear and less affected by noise. 
The case where no smearing is applied, is shifted up with respect
to others, however due to the small signal to noise ratio we could get only points in the range $r/a \in[1:6]$.
The plots with the fit curve are given in Fig. \ref{fig:pot-HYP}, where we show the fit for ``GEVP HYP(2) sm 12" case, 
{\cred where we applied a HYP (or HYP2) smearing step} to all gauge configurations and then solved the GEVP problem, described in the previous section, using a sHYP smearing in the spatial direction for the construction of the matrix $\hat{W}(r,t)$.
The fit function in this case is $V(r) = aV_{cont}(r) + \delta V(r)$, where $\delta V(r)$ is a correction due to the HYP(2) smearing, see Refs. \cite{hasenfratz2002static,hasenfratz2001flavor} for an explicit expression of this term. 
We did not attempt to fit the few obtained points in the no-smearing case, as the points at small $r/a$ can be affected by lattice artifacts, and thus an ansatz different than Eq. (\ref{eq:vr}) should be taken in this case.

From the fit results reported in Table \ref{tab:vr1} we can compare the string tension $\sigma$ and the Coulomb-parameter $\alpha$ for different smearing choices. 
We observe that although in the same ballpark, the results for different smearing choices show slight inconsistencies among each other, which can be explained with systematic effects that will not be addressed in this work. 
The separation of the two results for HYP and HYP2 smearing comes from an overall additive constant, as well as the correction term $\delta V(r)$ between HYP and HYP2 smearing, as discussed in Refs. \cite{hasenfratz2001flavor,Okiharu:2004tg,DellaMorte:2003mn}. 
Analyzing the signal to noise ratio and the $\chi^2$ of the fits, the use of 
``GEVP" smearing procedure seems to give better results and in this case the potential for large $r/a$ is better approximated by the continuum potential given in Eq. (\ref{eq:vr}), see Fig. \ref{fig:pot-HYP}.

\begin{table}[htb]
\centering
\begin{tabular}{ccccc}
Type &	$\alpha$ &	$\sigma[GeV^2]$	& $\chi^2/ndf$ & range $r/a$\\
\hline
HYP2+0		&0.346(7)	&0.267(3)	&1.01	&[2:11]\\
HYP2+sHYP sm 4	&0.372(53)	&0.256(13)	&0.97	&[2:11]\\
HYP2+sHYP sm 10	&0.430(42)	&0.241(9)	&1.08	&[2:11]\\
GEVP HYP2 sm 12	&0.445(10)	&0.235(2)	&1.04	&[4:12]\\
HYP2+APE 0.5	&0.346(8)	&0.267(4)	&0.65	&[4:14]\\
HYP2+APE 0.7	&0.346(9)	&0.267(4)	&1.54	&[3:11]\\
HYP+0	    	&0.318(7)	&0.291(3)	&1.31	&[4:12]\\
HYP+sHYP sm 4	& 0.368(97)	&0.268(27)	&0.28	&[2:12]\\
HYP+sHYP sm 10	&0.468(38)	&0.238(9)	&0.27	&[2:12]\\
GEVP HYP sm 12	&0.470(9)	&0.234(2)	&0.64	&[4:16]\\
HYP+APE 0.5	& 0.318(7)	&0.291(3)	&1.39	&[3:8]\\
HYP+APE 0.7	& 0.458(32)	&0.241(8)	&0.96	&[4:12]\\
\hline
\end{tabular}
\caption{Parameters: $\alpha$ and $\sigma$, of the fit function $V(r)$ in Eq. (\ref{eq:vr}) for different smearing choices.}
\label{tab:vr1}
\end{table}

Now we can also compute the Sommer parameter $r_0$ from the relation

\begin{equation}
r_0^2 F(r_0) = 1.65
\end{equation}

\noindent
where $F(r) = V'(r)$. 

\begin{figure}[htb]
\centerline{
\includegraphics[scale=1.3]{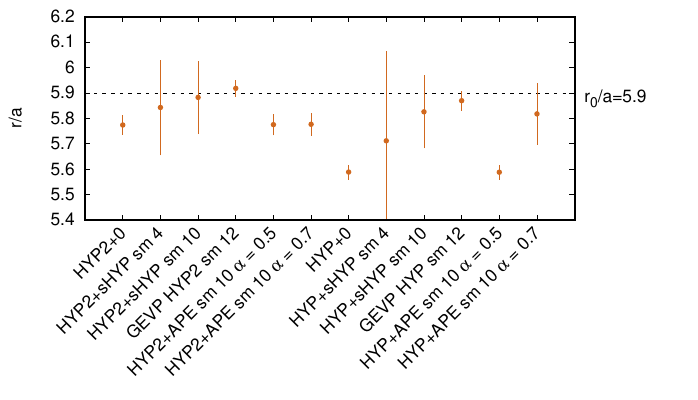}}
\caption{Sommer parameter $r_0/a$ for different smearing choices. It is compatible with $r_0/a =  5.9$, given in Ref. \cite{fritzsch2012strange}.}
\label{fig:pot-r0}
\end{figure}

This is an important crosscheck of the consistency of different smearing choices.
As we can observe in Fig. \ref{fig:pot-r0}, the Sommer parameter is consistent with {\cred the} value from the literature $r_0/a = 5.9$ \cite{fritzsch2012strange} in cases when either HYP and HYP2 smearing is applied; furthermore, the higher number of sHYP smearing steps we apply, the more precise {\cred a} result we obtain. On the other hand, when we apply no additional smearing in spatial direction (choices labeled as ``HYP2+0" and ``HYP+0") as well as in cases when ``APE" smearing is used instead of HYP, we obtain inconsistent results. It is important to note that the GEVP procedure already gives a good estimation of $r_0/a$ in combination with both HYP and HYP2 smearing.

\section{Conclusions and outlook}

We have reported on a preliminary study of static potentials between a quark-antiquark {\cred pair} in a full QCD simulation with $n_f=2$ {\cred based on Wilson loops}. Different choices how to smear gauge configurations combined with the GEVP procedure are deemed necessary to get reasonably good signals for our data.
From such static potentials we got the value of string tension and Sommer parameter and we compared them among different smearing procedures. 
This work is still very preliminary and further studies are important in order to get the remaining elements of the matrix of correlators in Eq. (\ref{eq:bali}) and then implement the Born-Oppenheimer approximation for the study of quarkonium states \cite{Bicudo:2019ymo,Prelovsek:2019ywc,Bicudo:2015vta,Bicudo:2016ooe}. 
We plan to explore additional techniques for noise reduction and extend the calculation for different gauge ensembles in order to study string breaking in the continuum limit. 

\section*{Acknowledgements}

This work is supported by Physik-IT at Ludwig Maximilian University of Munich. 
P.B. and N.C. thank CeFEMA, FCT contract UID/CTM/04540/2013.
N.C. acknowledges the FCT contract
SFRH/BPD/109443/2015.

\bibliographystyle{ieeetr}

\end{document}